\documentclass[twocolumn,showpacs,amsmath,amssymb,aps,pra,floatfix,nofootinbib,superscriptaddress]{revtex4}
\usepackage{graphicx,graphics,times,color}
\usepackage{bm}
\usepackage{longtable}
\usepackage{color}
\usepackage{lipsum}

\def\be{\begin{equation}}
\def\ee{\end{equation}}
\def\ba{\begin{eqnarray}}
\def\ea{\end{eqnarray}}
\def\la{\langle}
\def\ra{\rangle}

\begin{document}

\title{Comparing different modes of quantum state transfer in a XXZ spin chain}

\author{Sima Pouyandeh}
\affiliation{Department of Physics, Isfahan University of Technology, Isfahan 84156-83111, Iran}

\author{Farhad Shahbazi}
\affiliation{Department of Physics, Isfahan University of Technology, Isfahan 84156-83111, Iran}

\date{\today}

\begin{abstract}
We study the information transferring ability of a spin-$1/2$ XXZ Hamiltonian for two different modes of state transfer, namely, the well studied attaching scenario and the recently proposed measurement induced transport. The latter one has been inspired by recent achievements in optical lattice experiments for local addressability of individual atoms and their time evolution when only local rotations and measurements are available and local control of the Hamiltonian is very limited. We show that while measurement induced transport gives higher fidelity for quantum state transfer around the isotropic Heisenberg point, its superiority is less pronounced in non-interacting free fermionic XX phase. Moreover, we study the quality of state transfer in the presence of thermal fluctuations and environmental interactions and show that  measurement scheme gives higher fidelity for low temperatures and weak interaction with environment.

\end{abstract}

\pacs{03.67.-a, 03.67.Hk, 37.10.Jk, 32.80.Hd}

\maketitle

\section{Introduction} \label{sec1}

In the recent years the condensed matter many-body systems have extensively been used for realization of quantum information tasks such as quantum state transfer \cite{state-transfer-book}. In particular, finite many-body systems, such as spin chains, have been proposed to operate as data-bus \cite{bose03} through their natural time evolution. Very recently, experimental realizations of such ideas have become viable in Nuclear Magnetic Resonance (NMR) \cite{NMR-state-transfer}, coupled optical wave guides \cite{optical-waveguide-transfer} and optical lattices \cite{Bloch-spin-wave,Bloch-magnon}. However, there is a threshold between the quality of state transfer and the amount of control that one can have over the system. Perfect state transfer comes at its price \cite{Kay-review}. To achieve such task one has to engineer the spectrum of the system by either engineering all couplings \cite{christandl} or just the two at the boundaries \cite{bayat-gate}. Alternatively, one can also use precisely time controlled local or global fields \cite{state-transfer-time} for routing information or create resonant dynamics between the sender and receiver by using very weak couplings \cite{weak-couplings} or strong local magnetic fields \cite{strong-magnetic-field}. In all these methods, a very fine control over the couplings (either local or global) is needed which is often not available for the most physical systems used in laboratories. Thus, one may think of the minimal control of the most simplest systems (with uniform couplings) for achieving the highest possible fidelity. 
In a uniform XX chain one can achieve pretty good state transfer, with arbitrarily high fidelity, for very particular lengths though one has to wait for very long times \cite{yasser}.

The most studied mode of transmission so far is to attach a qubit encoding an \emph{unknown} quantum state to the spin chain which is motivated by the need to link quantum registers. This needs fine local control over a single bond of the Hamiltonian which has to be switched on and off at regular times which makes it very challenging for physical realizations. Moreover, for the case of anti-ferromagnetic systems this mode of transfer does not harness the inherent entanglement naturally existed in the ground state of such systems. The initial state of the channel changes the quality of transfer for both free fermions and interacting Hamiltonians \cite{bayat-initialization} and in the case of non-interacting fermionic systems it is possible to engineer the system such that it works fully independent of the initial state \cite{DiFranco-intialization}.

Recent achievements in optical lattice experiments for local addressability of atoms with the resolution of single sites \cite{Bloch-single-site-addressing} has opened a totally new window for single site unitary operations and measurements \cite{Mieschede-measurement}. The propagation of spin waves \cite{Bloch-spin-wave} and magnon bound states \cite{Bloch-magnon} have been observed in ferromagnetic spin chains realized by trapped atoms in optical lattices. Motivated by these achievements, new modes of transfer in many-body strongly correlated systems have been proposed which are not based on attaching scenarios which demand very fine control of a single bond in the Hamiltonian. In Ref.~\cite{bayat-densecoding} a single qubit unitary operation together with local measurements on a system initialized in a series of singlets are used for achieving more than one bit of information transfer in a single shot in the same spirit of quantum dense coding \cite{Bennet-densecoding}. We have also put forward a proposal for Measurement Induced Transport (MIT) \cite{a-measurement} without changing the Hamiltonian for information transfer which exploits only local measurement and unitary rotation of single qubits, the two ingredients both available in optical lattices.
In such scenario, a known quantum state is encoded in the first qubit of the system through a local measurement followed by a unitary operation and then the natural time evolution of the system transfers this quantum state to the receiver site. As the quantum state is assumed to be \emph{known} this mode of transfer can also be considered as remote quantum state preparation \cite{Bennet-remote}.

In this paper, by exploiting exact diagonalization we study the information transferring ability of a uniform spin-1/2 XXZ Hamiltonian throughout its anti ferromagnetic and {N{\'e}el} phases for two different modes of transfer namely: (i) the well studied attaching scenario and; (ii) the recently proposed MIT scheme. We also compare their performance in the presence of imperfections such as decoherence and thermal fluctuations. We show that while the two schemes give the same fidelity for non-interacting free fermionic systems the measurement scenario provides a sensibly better performance for the interacting systems. Moreover, the measurement procedure for state transfer shows more robustness against various imperfections such as thermal fluctuations.

The structure of the paper is as following. In section \ref{sec2} we introduce the model, in section \ref{sec3} the state transfer is investigated by calculation the average fidelity in both measurement and attaching scenarios and the results are compared. In section \ref{sec4} the effect of thermal fluctuations and decoherenc is studied. Finally, in section \ref{sec5} we summarize our results.

\begin{figure} \centering
\includegraphics[width=8cm,height=6cm,angle=0]{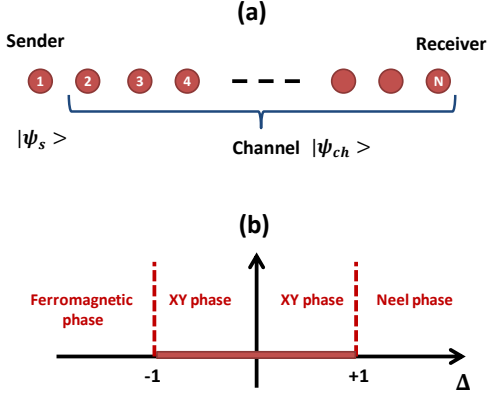}
\caption{(Color online) (a) An array of interacting spin 1/2 particles of a XXZ hamiltonian in which the first spin is initialized to a quantum state $|\psi_{s}\ra$ and the rest of the system is assumed to be in the quantum state $|\psi_{ch}\ra$. (b) Phase diagram of XXZ model for $J>0$. The parameter $\Delta$ is the anisotropy in the z direction.}
\label{fig1}
\end{figure}

\section{Spin Chain Quantum State Transfer} \label{sec2}

In this section, we review the most general scheme of quantum state transfer across spin chains. Consider a system of $N$ spin-1/2 interacting particles. The first qubit is initialized by the sender to a very general quantum state of
\begin{equation}\label{psi_s}
|\psi_{s}\ra=\cos{(\frac{\theta}{2})}|0\ra+e^{i\phi}\sin{(\frac{\theta}{2})}|1\ra
\end{equation}
where $|0\ra$ and $|1\ra$ are the two levels of the qubit and $0\leq \theta\leq \pi$ and $0\leq\phi\leq2\pi$ determine the location of this state on the surface of the Bloch sphere. The rest of the system (i.e. the qubits $2,3,...,N$) are called channel and they are assumed to be in a quantum state $|\psi_{ch}\ra$. A schematic picture of the situation is shown in Fig.~\ref{fig1}(a). The total quantum state of the system is written as
\begin{equation}\label{initial-psi}
|\Psi(0)\ra=|\psi_{s}\ra \otimes |\psi_{ch}\ra.
\end{equation}
At time $t=0$ the interaction between the qubits is switched on and the whole system evolves under the action of a Hamiltonian $H$. At any time $t$ the quantum state of the whole system is given by
\begin{equation}\label{psi_t}
|\Psi\ra=e^{-iHt} |\Psi(0)\ra.
\end{equation}
The last spin, namely site $N$, is assumed to be the receiver site which its density matrix at time $t$ can be computed as
\begin{equation}\label{rho}
\rho_{N}(t)=Tr_{\widehat{N}}|\Psi(t)\ra\langle\Psi(t)|
\end{equation}
where $ Tr_{\widehat{N}} $ means that we trace out all spins except the site $N$.
To quantify the quality of state transferring we calculate the fidelity of the last site and the initial state $|\psi_s\ra$
\begin{equation}\label{F}
F(\theta,\phi,t)=\langle\psi_s|\rho_{N}|\psi_s\ra.
\end{equation}
This fidelity depends on input parameters $\theta$ and $\phi$ and hence in order to have an input independent quantity we average over all possible pure input states on the surface of the Bloch sphere
\begin{equation}\label{Fav}
F_{av}(t)=\frac{1}{4\pi} \int_{\theta=0}^{\theta=\pi}\int_{\phi=0}^{\phi=2\pi}{F(\theta,\phi,t) \sin(\theta) d\theta d\phi}.
\end{equation}
This average fidelity can be simplified if we restrict ourselves to a certain class of Hamiltonians which conserve the number of excitations. Mathematically this means that $[H,S_{z}^t]=0$, where, $S_z^{t}$ is the total spin operator in the $z$ direction. A large class of Hamiltonians, including the XXZ, which naturally appears in nature have this property. Using the exact form of $|\psi_s\ra$, given in Eq.~(\ref{psi_s}), and using the conservation of excitations one can write the average fidelity as
\begin{widetext}
\begin{eqnarray} \label{Fav_t_0}
F_{av}(t)&=& \frac{1}{6} \left\{ \la\ 1,\psi_{ch}|e^{+iHt} P_{00}^{(N)} e^{-iHt} |1,\psi_{ch}\ra
+ \la0,\psi_{ch}|e^{+iHt} P_{11}^{(N)} e^{-iHt} |0,\psi_{ch}\ra \right\} \cr
&+& \frac{1}{3} \left\{ \la 0,\psi_{ch}|e^{+iHt} P_{00}^{(N)} e^{-iHt} |0,\psi_{ch}\ra
+ \la1,\psi_{ch}|e^{+iHt} P_{11}^{(N)} e^{-iHt} |1,\psi_{ch}\ra \right\} \cr
&+& \frac{1}{3} {\mathbf {abs}}\left\{ \la 1,\psi_{ch}|e^{+iHt} P_{10}^{(N)} e^{-iHt} |0,\psi_{ch} \ra \right\},
\end{eqnarray}
\end{widetext}
where, $P_{ij}^{(N)}=|i\ra \la j|$ is the single qubit projection operator acting on site $N$. As the formula for the average fidelity shows, to have a full characterization of the information transfer one has to only compute the evolution of two quantum states namely $|0,\psi_{ch}\ra$ and $|1,\psi_{ch}\ra$.

\section{The Phase Diagram of the XXZ Hamiltonian} \label{sec3}

In this paper we assume that the Hamiltonian $H$ is a XXZ Heisenberg type which can be written as
\begin{equation}\label{H}
H=J\sum_{k=1}^{N-1} (\sigma^{x}_{k} \sigma^{x}_{k+1}+\sigma^{y}_{k} \sigma^{y}_{k+1}+\Delta\sigma^{z}_{k}\sigma^{z}_{k+1})
\end{equation}
where $ \sigma^{x}_{k},\sigma^{y}_{k},\sigma^{z}_{k} $ are the Pauli operators acting on site $k$, the parameter $J$ is the exchange coupling and $\Delta$ is the anisotropy in the $z$ direction. This Hamiltonian has a rich phase diagram for $J>0$ \cite{XXZ-phase-diagram} as the anisotropy $\Delta$ varies which is schematically shown in Fig.~\ref{fig1}(b). For $\Delta<-1$, the system is ferromagnetic and its ground state is doubly degenerate with all spins aligned either in the $|0\ra$ or $|1\ra$ states. For $-1 \leq \Delta \leq 1$ the system enters a new gapless regime, called XY phase, in which the ground state is unique and includes the free fermionic XX Hamiltonian (i.e. $\Delta=0$) and the fully isotropic Heisenberg one (i.e. $\Delta=1$).
 The $\Delta >1$ is called the Neel phase with a gapped spectrum and nonzero staggered magnetization. In this paper we discuss the quantum state transferring for $\Delta>0$, which is the more relevant case in terms of experimental realization. 

\section{Modes of Transport: Attaching versus MIT} \label{sec4}

In the mode of attaching for quantum state transfer across a spin chain the quantum state $|\psi_{ch}\ra$ is the ground state of the channel (qubits $2,3,...,N$) while the first qubit is decoupled and is independently initialized to $|\psi_s\ra$. Then by switching on the interaction between the qubit 1 and the rest of the system the evolution starts as the state $|\Psi(0)\ra$, given in Eq.~(\ref{initial-psi}), is not an eigenstate of the total Hamiltonian $H$.

On the other hand, in the mode of MIT \cite{a-measurement} for state transfer there is no control on the Hamiltonian and it does not change during the process. Initially the whole system (qubits $1,2,...,N$) is prepared in its ground state $|GS\ra$. To encode the quantum state $|\psi_s\ra$ one measures the first qubit of the chain in the eigenstates of the $\sigma_z$ operator which the outcome of measurement will be either $|0\ra$ or $|1\ra$ with probability of 1/2 for each. According to the outcome of measurement the quantum state of the whole system collapses as
\begin{eqnarray} \label{psi_collapse}
|0\ra: |GS\ra \rightarrow |0,\Phi_0\ra \cr
|1\ra: |GS\ra \rightarrow |1,\Phi_1\ra
\end{eqnarray}
where $|\Phi_0\ra$ and $|\Phi_1\ra$ are the quantum states of the channel (i.e. qubits $2,3,...,N$) when the first qubit is collapsed to $|0\ra$ and $|1\ra$ respectively. To finalize the initialization of the system into a quantum state of the type of Eq.~(\ref{psi_s}) one has to apply a further unitary operator $R_0$ or $R_1$, according to the outcome of measurement, on the first qubit to rotate its quantum state into $|\psi_s\ra$. One can easily determine the form of $R_0$ and $R_1$
\begin{eqnarray} \label{R1_R2}
R_0 &=& |\psi_{s}\ra \la 0 | + |\widetilde{\psi}_{s}\ra \la 1 | \cr
R_1 &=& |\widetilde{\psi}_s\ra \la 0 | + |\psi_{s}\ra \la 1 |
\end{eqnarray}
Where $|\widetilde{\psi}_{s}\ra$ is the orthogonal counterpart of $|\psi_s\ra$, i.e $\la \psi_s|\widetilde{\psi}_s\ra=0 $, given by
\begin{equation}\label{psi_s_ortho}
|\widetilde{\psi}_{s}\ra=\sin{(\frac{\theta}{2})}|0\ra-e^{i\phi}\cos{(\frac{\theta}{2})}|1\ra.
\end{equation}
At this stage the initialization of the system is accomplished as the state of the system takes the form of Eq.~(\ref{initial-psi}) for both measurement outcomes
\begin{eqnarray} \label{psi_0_measure}
|0\ra: |\Psi_0(0)\ra&=& |\psi_{s}\ra \otimes |\Phi_0\ra \cr
|1\ra: |\Psi_1(0)\ra&=& |\psi_{s}\ra \otimes |\Phi_1\ra.
\end{eqnarray}
Neither $|\Psi_0(0)\ra$ nor $|\Psi_1(0)\ra$ are the eigenvectors of the Hamiltonian and thus evolve under the action of the XXZ Hamiltonian $H$ as given in Eq.~(\ref{psi_t}). For each measurement outcome, the final average fidelity is similar to $F_{av}(t)$ in Eq.~(\ref{Fav_t_0}) by replacing $|\psi_{ch}\ra$ with $|\Phi_0\ra$ or $|\Phi_1\ra$ accordingly.

\begin{figure} \centering
\includegraphics[width=9cm,height=7cm,angle=0]{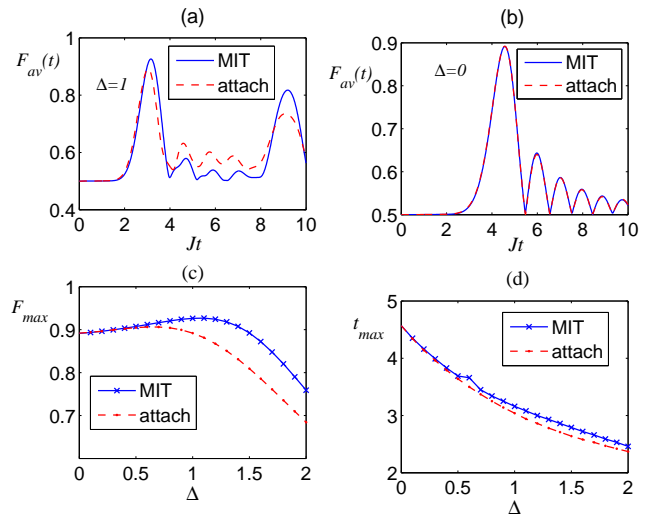}
\caption{(Color online) (a) Average fidelity in terms of time for $ \Delta=1 $ in a chain of length $ N=16 $ for both measurement and attaching scenarios. (b) $ F_{av}(t) $ versus time for $ \Delta=0$ and $ N=16 $ (c)The average fidelity at its first peak versus $\Delta$ for $\Delta>0$ and $N=16$. (d) The optimal time, when $F_{av}$ peaks for the first, time for $N=16$. }
\label{fig2}
\end{figure}

In Figs.~\ref{fig2}(a) and (b) the $F_{av}(t)$ is plotted versus time for both scenarios in a chain of length $N=16$ for $\Delta=1$ and $\Delta=0$ respectively. The average fidelity $F_{av}(t)$ starts evolving after a initial pause needed for information to reach the last site. As it is clear from these figures, the measurement induced transport gives a better quality for the case of isotropic Heisenberg Hamiltonian (i.e. $\Delta=1$) while for the XX Hamiltonian (i.e. $\Delta=0$) the two scenarios are identical. This is indeed due to the non-interacting fermionic nature of the XX Hamiltonian which makes the attainable average fidelity almost independent of the initialization mechanism \cite{bayat-initialization}. To have a more concrete comparison between the two scenarios, in Fig.~\ref{fig2}(c) we plot $ F_{max} $ versus $\Delta>0$ for a chain of length $N=16$. As this figure shows, in MIT strategy the anisotropic point $\Delta=1.1$ in the N{\'e}el phase gives the highest fidelity for a chain of length $N=16$. Our analysis shows that as the length of the chain becomes shorter, this point moves toward the XY phase. In attaching scenario, It is the point $\Delta=0.8$ in the XY phase that has the highest fidelity for a chain of $N=16$. Moreover, for longer chains this point gets closer to the isotropic point $\Delta=1$.
In Fig.~\ref{fig2}(d) the $t_{max}$ is plotted in terms of $\Delta$,  illustrating  that the optimal time is almost equal for both strategies and by increasing $\Delta$ the evolution becomes faster, which could be due to the increasing of energy gap in the finite size system. The results shown in Figs.~\ref{fig2}(c) and (d) are fully consistent with the ones for entanglement distribution across the XXZ Hamiltonian reported for the attaching scenario in Ref.~\cite{bayat-xxz}.

\begin{figure} \centering
\includegraphics[width=9cm,height=4cm,angle=0]{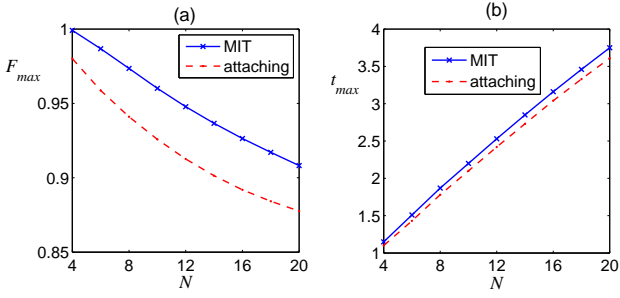}
\caption{(Color online) (a) $F_{av}$ in its first peak in terms of N for $ \Delta=1.1$ and both scenarios. (b) The Optimal time when $F_{av}$ peaks for the first time as a function of N for $ \Delta=1.1 $. }
\label{fig3}
\end{figure}

To compare the scalability of the two scenarios, in Fig.~\ref{fig3}(a) we plot the $F_{max}$ as a function of length $N$ in the XXZ chain with $\Delta=1.1$ for both attaching and MIT mechanisms. As this figure clearly shows the measurement induced dynamics gives a higher fidelity which even becomes more prominent for longer chains. In Fig.~\ref{fig3}(b) the optimal time $t_{max}$ is depicted versus $N$ for both scenarios. As expected the $t_{max}$ grows linearly by increasing $N$ and the MIT is slightly slower in speed in compare to the attaching procedure.

\section{IMPERFECTIONS} \label{sec5}

\begin{figure*}
\includegraphics[width=13cm,height=6cm,angle=0]{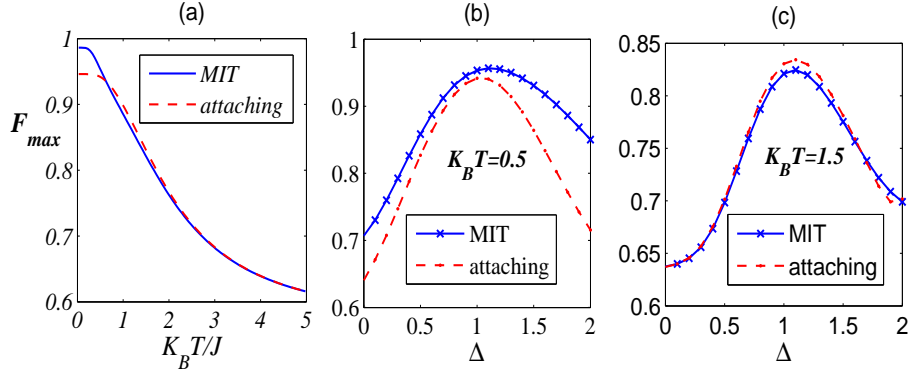}
\caption{(Color online) (a) $F_{max}$ in terms of temperature for isotropic case $ \Delta=1$ and a chain of length $N=10$ when in MIT the outcome of the measurement is $|0\ra$. (b) $ F_{max}$ at $ K_BT=0.5 $ and for $ N=10 $ (c) $ F_{max} $ versus $ \Delta $ at $ K_BT=1.5 $. }
\label{fig4}
\end{figure*}

No physical system is in zero temperature and thus any quantum system is inevitably in a thermal mixed state. Let's assume that the whole system is initialized in
\begin{equation}\label{rho_thermal}
\rho(0)=|\Psi_{s}\ra\langle\Psi_{s}| \otimes \rho_{ch}
\end{equation}
where $\rho_{ch}$ is a mixed state and not necessarily a thermal state. The quantum state at an arbitrary time $t$ is
\begin{equation}\label{rho_t}
\rho(t)=e^{-iHt}\rho(0)e^{+iHt}.
\end{equation}
Just as explained above one can trace out all spins except the last one (i.e. site $N$) and compute the average fidelity which becomes
\begin{widetext}
\begin{eqnarray} \label{Fav_thermal_t}
F_{av}(t)&=& \frac{1}{6} \left\{ Tr\{ e^{-iHt} ( |1\ra \la 1| \otimes \rho_{ch}) e^{+iHt} P_{00}^{(N)} \}
+ Tr\{ e^{-iHt} ( |0\ra \la 0| \otimes \rho_{ch}) e^{+iHt} P_{11}^{(N)} \} \right\} \cr
&+& \frac{1}{3} \left\{ Tr\{ e^{-iHt} ( |0\ra \la 0| \otimes \rho_{ch}) e^{+iHt} P_{00}^{(N)} \}
+ Tr\{ e^{-iHt} ( |1\ra \la 1| \otimes \rho_{ch}) e^{+iHt} P_{11}^{(N)} \right\} \cr
&+& \frac{1}{3} {\mathbf {abs}}\left\{ Tr\{ e^{-iHt} ( |0\ra \la 1| \otimes \rho_{ch}) e^{+iHt} P_{01}^{(N)} \right\}.
\end{eqnarray}
\end{widetext}

In the attaching scenario, the channel itself is in a thermal state of the form $\frac{e^{-\beta H_{ch}}}{Z}$, where, $\beta=1/K_{B}T$ ($T$ is temperature and $K_{B}$ is the Boltzmann constant). Hence, by replacing $\rho_{ch}$ in the above formula by $\frac{e^{-\beta H_{ch}}}{Z}$ one computes the average fidelity of the attaching scenario for the thermal states.

On the other hand for MIT we assume that the whole system is initially in a thermal state as $\frac{e^{-\beta H}}{Z}$ and then the first qubit is measured and according to the outcome of measurement the quantum state of the system collapses to
\begin{eqnarray} \label{rho_0_measure}
|0\ra&:& \frac{e^{-\beta H}}{Z} \rightarrow |0\ra \la 0| \otimes \rho_{0} \cr
|1\ra&:& \frac{e^{-\beta H}}{Z} \rightarrow |1\ra \la 1| \otimes \rho_{1}.
\end{eqnarray}
Then by applying the operator $R_0$ or $R_1$ on the first qubit the initialization is completed and the quantum state of the form of Eq.~(\ref{rho_thermal}) is formed
\begin{eqnarray} \label{rho_0_measure}
|0\ra&:& |\psi_s\ra \la \psi_s| \otimes \rho_{0} \cr
|1\ra&:& |\psi_s\ra \la \psi_s| \otimes \rho_{1}.
\end{eqnarray}
Now one can replace $\rho_{ch}$ in Eq.~(\ref{Fav_thermal_t}) by either $\rho_0$ or $\rho_1$ to compute the average fidelity of MIT scenario according to each outcome of the measurement.

In Fig.~\ref{fig4}(a), the average fidelity at its maximum, $F_{max}$ is plotted as a function of temperature for isotropic point $\Delta=1$ and for a chain of $ N=10$. As figure shows, the average fidelity decreases by increasing the temperature for both MIT and attaching schemes. In low temperatures, there exist a plateau over which the average fidelity remains constant and its width determines the thermal stability of the system. As it is evident from the figure, the plateau is smaller for MIT but since the fidelity is higher it gives a better performance for low temperatures. In higher temperatures both MIT and attaching scenarios are almost the same. we also found that the optimal time $t_{max}$ at which the average fidelity peaks is almost independent of temperature, consistent with Ref.~\cite{bayat-thermal}. In Fig.~\ref{fig4}(b) and (c) we plot $ F_{max} $ versus $\Delta$ for two different temperatures $ K_BT=0.5 $ , $ K_BT=1.5 $. It is seen that the anisotropic point of $ \Delta=1.1 $ and the isotropic point  $\Delta=1$ are the best points for MIT and attaching scenarios, respectively.

\begin{figure}
\includegraphics[width=8cm,height=7cm,angle=0]{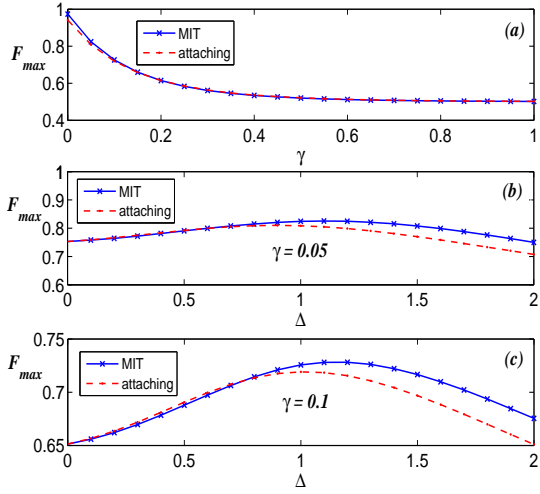}
\caption{(Color online) (a) $F_{max}$ as a function of $\gamma$ for $\Delta=1$ and a chain of length $N=8$ when in MIT the outcome of the measurement is $|0\ra$ (b) $ F_{max} $ in terms of $\Delta$ at $ \gamma=0.05 $ and for $ N=8$ (c)  $F_{max}$ versus $\Delta$ at $\gamma=0.1$. }
\label{fig5}
\end{figure}

Another imperfection that we consider is decoherency, as in real physical situations the system is not isolated perfectly. Here, we investigate the interaction of the system with its environment via a Lindblad equation in which
\begin{equation}\label{rho_dot_Lindblad}
      \dot{\rho}=-i [H,\rho]+l(\rho)
\end{equation}
In this equation $\l(\rho)$ is the Markovian evolution of the state $\rho$ and depends on the kind of interaction that system has with its environment.
If we assume an environment with no preferred direction, we should consider the interaction as
\begin{equation}\label{l_rho_sss}
       l(\rho)=-\gamma/3 \sum_{k=1}^{N}\sum_{\alpha}(\rho_{ij}-\sigma_{k}^{\alpha}\rho_{ij}\sigma_{k}^{\alpha})
\end{equation}
In these equations $\gamma$ is the coupling with environment and $\alpha$ gets $x$ , $y$ , $z$.

Let's consider the initial state of the system as
\begin{equation}\label{rho_Lindblad}
\rho(0)=|\Psi_{s}\ra\langle\Psi_{s}| \otimes \rho_{ch}
\end{equation}
where $\rho_s=|\psi_s\ra\la\psi_s|$ is the state that is going to be sent and $\rho_{ch}$ is the state of channel. As before we compute the average fidelity that with a little bit of maths becomes
\begin{eqnarray} \label{Fav_Lindblad_t}
F_{av}(t)&=& \frac{1}{6} \left\{ Tr (\Omega_{11}P_{00}
+ Tr(\Omega_{00}P_{11}) \right\} \cr
&+& \frac{1}{3} \left\{ Tr(\Omega_{00}P_{00})
+ Tr(\Omega_{11}P_{11}) \right\} \cr
&+& \frac{1}{3} {\mathbf {abs}}\left\{ Tr(\Omega_{01}P_{10}) \right\},
\end{eqnarray}
where
\begin{equation}\label{omega}
\Omega_{ij}=|i\ra \la j| \otimes \rho_{ch}
\end{equation}

The time evolution, is given by the Lindblad equation of Eq.~(\ref{rho_dot_Lindblad}), using Rung-Kutta method. In Fig.~\ref{fig5}(a) We plot the fidelity $F_{max}$ in terms of $\gamma$ in a chain of length $N=8$ and $\Delta=1$ using both MIT and attaching scenarios. As it is seen, fidelity goes down by increasing $\gamma$ very quickly for both schemes. However, MIT gives higher fidelity for small values of $\gamma$, and this improvement becomes more evident in longer chains indicating  the superiority of the MIT strategy over attaching in noisy environments. Figs.~\ref{fig5}(b) and (c) show $F_{max}$ as a function of $\Delta$ for $\gamma=0.05$ and $\gamma=0.1$. As it is seen, here the best point for quantum state transferring is again an anisotropic point $\Delta=1.1$ in MIT scenario and the isotropic point $\Delta=1$ for attaching scheme. Moreover, the superiority of the MIT over the attaching scenario is more evident in weak noises and it becomes more profound by increasing the length (due to numerical limitations we have data up to $N=10$) which shows that in long chains the MIT approach will be more effective. In the limit of high noise regime the superiority of the MIT goes away and the two methods give the same performance.

\section{Conclusion} \label{sec5}
In this paper we investigated two different mechanisms, namely MIT and attaching scenario for the quantum state transfer through a $S=1/2$ XXZ spin chain. It is shown that the MIT scenario gives higher fidelity in particular for $\Delta \geq 1$, while both scenarios have almost the similar speed. In the other points of the XY phase (in particular around the non-interacting fermionic point $\Delta=0$) the two schemes give almost the same results. The superiority of the MIT becomes even more evident for longer chains. Our analysis also shows that the best point for quantum state transferring depends on the length of the chain. For longer chains it occurs in the N{\'e}el phase while for shorter chains it moves towards the XY phase. In the presence of thermal fluctuations, MIT and attaching scenarios significantly differ at very low temperatures, while the attaching  mechanism shows more robustness against increasing of temperature. In the case of interaction with environment and for the isotropic decoherence,  the best point in the phase diagram is the anisotropic point $\Delta=1.1$ in the N{\'e}el phase for MIT strategy and the isotropic Heisenberg Hamiltonian with $\Delta=1$ for attaching scenario. When the decoherence is very strong both the MIT and attaching schemes give almost the same performance and the superiority of the MIT disappears.

{\em Acknowledgements:-} The authors are pleased to warmly thank Abolfazl Bayat and Vahid Karimipour for useful comments and  discussions.


\begin{thebibliography}{}

\bibitem{state-transfer-book} G. M. Nikolopoulos, Igor Jex, {\em Quantum State Transfer and Network Engineering}, Springer (2013); S. Bose, Contemporary Physics
{\bf 48}, 13 (2007).

\bibitem{bose03} S. Bose, Phys. Rev. Lett. {\bf 91}, 207901 (2003).

\bibitem{NMR-state-transfer} K. R. Koteswara Rao, T. S. Mahesh, A. Kumar, arXiv:1307.5220.

\bibitem{optical-waveguide-transfer} A. Perez-Leija, R. Keil, A. Kay, H. Moya-Cessa, S. Nolte, L. C. Kwek, B. M. Rodríguez-Lara, A. Szameit, and D. N. Christodoulides, Phys. Rev. A {\bf 87}, 012309 (2013); M. Bellec, G. M. Nikolopoulos, and S. Tzortzakis, Optics Letters {\bf 37}, 4504 (2012).

\bibitem{Bloch-spin-wave} T. Fukuhara, {\em et al.}, Nature Phys. {\bf 9}, 235 (2013).

\bibitem{Bloch-magnon} T. Fukuhara, {\em et al.}, Nature {\bf 502}, 76 (2013).

\bibitem{Kay-review} A. Kay, Int. J. Quantum Inf. {\bf 8}, 641 (2010).

\bibitem{christandl} M. Christandl, N. Datta, A. Ekert, and A. J. Landahl, Phys. Rev. Lett. {\bf 92}, 187902 (2004);
C. Albanese, M. Christandl, N. Datta, and A. Ekert, Phys. Rev. Lett. {\bf 93}, 230502 (2004);
M. Christandl, N. Datta, T. C. Dorlas, A. Ekert, A. Kay, and A. J. Landahl, Phys. Rev. A {\bf 71}, 032312 (2005).

\bibitem{bayat-gate} L. Banchi, A. Bayat, P. Verrucchi, S. Bose, Phys. Rev. Lett. {\bf 106}, 140501 (2011).

\bibitem{state-transfer-time} F. Galve, D. Zueco, S. Kohler, E. Lutz, and P. H¨anggi, Phys.
Rev. A {\bf 79}, 032332 (2009);
D. Burgarth, V. Giovannetti, S. Bose, Phys. Rev. A {\bf 75}, 062327 (2007);
X. Wang, A. Bayat, S. Bose, S. Schirmer, Phys. Rev. A {\bf 82}, 012330 (2010).

\bibitem{weak-couplings} A. Wojcik, T. Luczak, P. Kurzynski, A. Grudka, T. Gdala, and M. Bednarska, Phys. Rev. A {\bf 72}, 034303 (2005);
A. Wojcik, T. Luczak, P. Kurzynski, A. Grudka, T. Gdala, and M. Bednarska, Phys. Rev. A {\bf 75}, 022330 (2007);
M. J. Hartmann, M. E. Reuter, and M. B. Plenio, New J. Phys. {\bf 8}, 94 (2006);
L. Campos Venuti, C. Degli Esposti Boschi, and M. Roncaglia, Phys. Rev. Lett. {\bf 99}, 060401 (2007);
L. Campos Venuti, S. M. Giampaolo, F. Illuminati, and P. Zanardi, Phys. Rev. A {\bf 76}, 052328 (2007);
G. Gualdi, S. M. Giampaolo, and F. Illuminati, Phys. Rev. Lett. {\bf 106}, 050501 (2011);
S. Paganelli, S. Lorenzo, T. J. G. Apollaro, F. Plastina, G. L. Giorgi, Phys. Rev. A {\bf 87}, 062309 (2013).

\bibitem{strong-magnetic-field} S. Lorenzo, T. J. G. Apollaro, A. Sindona, F. Plastina, Phys.
Rev. A {\bf 87}, 042313 (2013); K. Korzekwa, P. Machnikowski, P.
Horodecki, . arXiv:1403.7359.

\bibitem{yasser} C. Godsil, S. Kirkland, S. Severini, and J. Smith, Phys. Rev. Lett. 109, 050502 (2012); R. Sousa, Y. Omar, arXiv:1405.1296.

\bibitem{bayat-initialization} A. Bayat, L. Banchi, S. Bose, P. Verrucchi, Phys. Rev. A {\bf 83}, 062328 (2011).

\bibitem{DiFranco-intialization} C. Di Franco, M. Paternostro, M. S. Kim, Phys. Rev. Lett. {\bf 101}, 230502 (2008).

\bibitem{Bloch-single-site-addressing} J. F. Sherson, {\em et al.}, Nature {\bf 467}, 68 (2010); C. Weitenberg, {\em et al.}, Nature {\bf 471}, 319 (2011).

\bibitem{Mieschede-measurement} M. Karski, {\em et al.}, New J. Phys. {\bf 12}, 065027 (2010).

\bibitem{bayat-densecoding} S. Yang, A. Bayat, S. Bose, Phys. Rev. A {\bf 84}, 020302(R) (2011).

\bibitem{Bennet-densecoding} C. H. Bennett, and S. J. Wiesner, Phys. Rev. Lett. {\bf 69}, 2881 (1992).

\bibitem{a-measurement} S. Pouyandeh, F. shahbazi, A. Bayat, arXiv:1403.3903.

\bibitem{Bennet-remote} C. H. Bennett, et al., Phys. Rev. Lett. {\bf 87}, 077902 (2001).

\bibitem{XXZ-phase-diagram}. H. Mikeska, A. Kolezhuk, Lecture Notes in Physics {\bf 645}, pp. 1-83 (2004).

\bibitem{bayat-xxz} A. Bayat and S. Bose, Phys. Rev. A {\bf 81}, 012304 (2010).

\bibitem{bayat-thermal} A. Bayat, V. Karimipour, Phys. Rev. A, {\bf 71}, 042330 (2005).

\end{thebibliography}
\end{document}